%% file: main.tex
\documentclass[12pt]{article}
\pdfoutput=1
\usepackage[utf8x]{inputenc}      
\usepackage{hyperref}
\usepackage{amsmath,amssymb,mathtools}
\usepackage[T1]{fontenc}          
\usepackage{booktabs,tabularx}
\usepackage{graphicx,subfig}
\usepackage{xspace}
\usepackage[usenames]{xcolor}\definecolor{fscolor}{RGB}{44,118,255}
\usepackage{geometry}
\usepackage{todonotes}
\usepackage{listings}
\usepackage[absolute]{textpos}
\usepackage[many]{tcolorbox}
\usepackage{xparse}
\usepackage[font=small,labelfont=bf,format=plain,margin=0.05\textwidth]{caption}
\usepackage{bbm}
\usepackage{tabularx}
\usepackage{cite}
\usepackage{soul}
\usepackage{afterpage}

\allowdisplaybreaks

\evensidemargin \oddsidemargin
\input{abbr.tex}

\begin{document}

\thispagestyle{empty}
\def\thefootnote{\fnsymbol{footnote}}

\begin{flushright}
CP3-Origins-2020-12 DNRF90 \\
DESY 20-155\\
FR-PHENO-2020-013
\end{flushright}
\vspace{3em}
\begin{center}
{\Large\bf Hybrid calculation of the MSSM Higgs boson masses using the complex THDM as EFT}
\\
\vspace{3em}
{
Henning Bahl$^a$\footnote{email: henning.bahl@desy.de},
Nick Murphy$^b$\footnote{email: murphy@cp3.sdu.dk},
Heidi Rzehak$^{b,\, c}$\footnote{email: heidi.rzehak@physik.uni-freiburg.de}
}\\[2em]
{\sl ${}^a$Deutsches Elektronen-Synchrotron DESY, Notkestra{\ss}e 85, 22607 Hamburg, Germany.}
{\sl ${}^b$CP3-Origins, University of Southern Denmark, Campusvej 55, 5230 Odense M, Denmark.}
{\sl ${}^c$Albert-Ludwigs-Universit\"at Freiburg, Physikalisches Institut, Hermann-Herder-Stra\ss e 3, 79104 Freiburg, Germany.}
\def\thefootnote{\arabic{footnote}}
\setcounter{page}{0}
\setcounter{footnote}{0}
\end{center}
\vspace{2ex}
\begin{abstract}
{}

\input{00_abstract.tex}

\end{abstract}

\newpage
\tableofcontents
\newpage
\def\thefootnote{\arabic{footnote}}


\section{Introduction}
\label{sec:01_intro}

\input{01_intro.tex}


\section{EFT calculation}
\label{sec:02_EFTimprovements}

\input{02_EFTimprovements.tex}


\section{Combination with fixed-order calculation}
\label{sec:03_FOcomb}

\input{03_FOcomb.tex}


\section{Numerical results}
\label{sec:04_results}

\input{04_results.tex}


\section{Conclusions}
\label{sec:05_conclusions}

\input{05_conclusions.tex}


\section*{Acknowledgments}
\sloppy{
We thank Ivan Sobolev and Georg Weiglein for useful discussions as well as Stefan Liebler for help with \texttt{SusHi}. H.B. is supported by the Deutsche Forschungsgemeinschaft (DFG, German Research Foundation) under Germany‘s Excellence Strategy -- EXC 2121 ``Quantum Universe'' – 390833306. H.R.'s and N.M.'s work was partially funded by the Danish National Research Foundation, grant number DNRF90. H.R. is also supported by the German Federal Ministry for Education and Research (BMBF) under contract no.\ 05H18VFCA1. Part of this work was supported by a STSM Grant from COST Action CA16201 PARTICLEFACE.
}


\appendix


\section{Threshold corrections: explicit expressions}
\label{app:06_thresholds}

\input{06_app_thresholds.tex}



\newpage

\bibliographystyle{JHEP.bst}
\bibliography{bibliography}{}

\end{document}

%% file: abbr.tex
\newcommand{\mf}{\widehat \mu}

\newcommand{\sbe}{s_\beta}
\newcommand{\cbe}{c_\beta}
\newcommand{\tbe}{t_\beta}
\newcommand{\sbb}{s_\beta^2}
\newcommand{\cbb}{c_\beta^2}
\newcommand{\tbb}{t_\beta^2}

\newcommand{\CP}{\ensuremath{\mathcal{CP}}\xspace}
\newcommand{\DR}{{\ensuremath{\overline{\text{DR}}}}\xspace}
\newcommand{\MS}{{\ensuremath{\overline{\text{MS}}}}\xspace}

\newcommand{\OS}{{\text{OS}}\xspace}

\newcommand{\SUSY}{{\text{SUSY}}}

\newcommand{\Pdd}{\Phi_1^\dagger\Phi_1}
\newcommand{\Puu}{\Phi_2^\dagger\Phi_2}
\newcommand{\Pdu}{\Phi_1^\dagger\Phi_2}
\newcommand{\Pud}{\Phi_2^\dagger\Phi_1}

\newcommand{\FH}{\mbox{{\tt FeynHiggs}}\xspace}

\newcommand{\Fig}[1]{Fig.~\ref{#1}}
\newcommand{\Sec}[1]{Section~\ref{#1}}
\newcommand{\App}[1]{App.~\ref{#1}}
\newcommand{\Eq}[1]{Eq.~(\ref{#1})}

\newcommand{\Eqss}[2]{Eqs.~(\ref{#1})-(\ref{#2})}

\renewcommand{\Re}{\text{Re}}
\renewcommand{\Im}{\text{Im}}

\newcommand{\Mh}{\ensuremath{M_{h_1}}\xspace}

\newcommand{\MHp}{\ensuremath{M_{\Hpm}}\xspace}

\newcommand{\cp}{\ensuremath{{\cal CP}}}

\newcommand{\msusy}{\ensuremath{M_\SUSY}\xspace}

\newcommand{\xt}{\ensuremath{\widehat X_t}\xspace}

\newcommand{\at}{\ensuremath{\widehat A_t}\xspace}

\newcommand{\tev}{\,\, \mathrm{TeV}}
\newcommand{\gev}{\,\, \mathrm{GeV}}

\newcommand{\order}[1]{\ensuremath{{\cal O}(#1)}}
\newcommand{\al}{\alpha}
\newcommand{\als}{\al_s}
\newcommand{\alt}{\al_t}
\newcommand{\alb}{\al_b}

\newcommand{\Hpm}{\ensuremath{H^\pm}\xspace}

\newcommand{\pMi}{\ensuremath{\phi_{M_1}}\xspace}
\newcommand{\pMii}{\ensuremath{\phi_{M_2}}\xspace}
\newcommand{\pMiii}{\ensuremath{\phi_{M_3}}\xspace}
\newcommand{\pMue}{\ensuremath{\phi_{\mu}}\xspace}

\newcommand{\pAt}{\ensuremath{\phi_{A_t}}\xspace}

\newcommand{\li}{{\lambda_1}}
\newcommand{\lii}{{\lambda_2}}
\newcommand{\liii}{{\lambda_3}}
\newcommand{\liv}{{\lambda_4}}
\newcommand{\lv}{{\lambda_5}}
\newcommand{\lvRe}{{\Re\lambda_5}}
\newcommand{\lvIm}{{\Im\lambda_5}}
\newcommand{\lvi}{{\lambda_6}}
\newcommand{\lviRe}{{\Re\lambda_6}}
\newcommand{\lviIm}{{\Im\lambda_6}}
\newcommand{\lvii}{{\lambda_7}}
\newcommand{\lviiRe}{{\Re\lambda_7}}
\newcommand{\lviiIm}{{\Im\lambda_7}}

%% file: 00_abstract.tex
Recently, the Higgs boson masses in the Minimal Supersymmetric Standard Model (MSSM) and their mixing have been calculated using the complex Two-Higgs-Doublet Model (cTHDM) as an effective field theory (EFT) of the MSSM. Here, we discuss the implementation of this calculation, which we improve in several aspects, into the hybrid framework of \FH by combing the cTHDM-EFT calculation with the existing fixed-order calculation. This combination allows accurate predictions also in the intermediate regime where some SUSY particles are relatively light, some relatively heavy and some in between. Moreover, the implementation provides precise predictions for the Higgs decay rates and production cross-sections.

%% file: 01_intro.tex
The measurements of an increasing number of SM-like observables by the experiments at the Large Hadron Collider (LHC) approach an accuracy that allows extensions of the Standard Model (SM) to be probed thoroughly beyond the search of yet undiscovered particles. Particularly important in this context are the measurements of the properties of the discovered Higgs boson: With the increased accuracy, these properties are turning into new precision observables that permit tests of many extensions of the Standard Model, namely the ones requiring changes in the Higgs sector. These extensions address questions that the SM cannot answer, such as what dark matter is made up of and why there is more matter than antimatter in the universe.

The results of the measurements performed at the LHC experiments confirm so far the predictions of the SM and therefore constrain the parameter space of possible SM extensions to the parts, which reproduce the SM-like behaviour.

One such model, which can, for example, provide a dark-matter candidate as well as additional sources of \CP violation needed for the explanation of the matter--antimatter asymmetry, is the Mininmal Supersymmetric Standard Model (MSSM), built upon the concept of Supersymmetry (SUSY). It extends the SM by adding a second Higgs doublet, turning the SM into a Two-Higgs-Doublet Model (THDM), as well as a superpartner to each THDM degree of freedom. The two Higgs doublets lead to five physical Higgs bosons: At tree-level, these are the \CP-even $h$ and $H$ bosons, the \CP-odd $A$ boson and the charged \Hpm bosons. Beyond the tree-level, quantum corrections lead to mixing of the tree-level mass eigenstates. In presence of \CP violation the true mass eigenstates of the neutral Higgs bosons are labelled by $h_{1,2,3}$ and are in general mixtures of \cp-even as well as \cp-odd states.

In the context of the MSSM, especially the Higgs boson mass measurement~\cite{Aad:2012tfa,Chatrchyan:2012xdj,Aad:2015zhl} is a unique opportunity to constrain the MSSM Higgs sector. While the Higgs boson mass is a free parameter in the SM, the mass of the MSSM SM-like Higgs boson can be predicted in terms of the model parameters.

In order to provide a precise prediction, the calculation of higher-order corrections is mandatory: Higher-order corrections are essential in the context of the MSSM, since they lead to a circumvention of the upper tree-level mass bound of the lightest MSSM Higgs boson, which is given by the $Z$-boson mass, $M_Z \approx 91$ GeV,  and thereby to an experimentally allowed Higgs-boson mass.\footnote{The scenario where a heavy MSSM Higgs boson plays the role of the discovered one is strongly constrained~\cite{Bahl:2018zmf}.} These higher-order corrections can also change the coupling behaviour of the Higgs bosons, for example by mixing \cp-even and \cp-odd states, and are therefore also important for predicting partial decay widths and production cross sections.

Given the importance of these quantum corrections, much work has been invested in calculating them.
These calculations fall into three categories:
\begin{itemize}
\item In the fixed-order (FO) approach, contributions to the Higgs two-point-vertex functions are calculated at a fixed-order of the couplings taking into account the complete model. This results in corrections of the Higgs-boson masses and mixings (see~\cite{Borowka:2014wla,Degrassi:2014pfa,Borowka:2015ura,Goodsell:2016udb,Passehr:2017ufr,Harlander:2017kuc,Borowka:2018anu,R.:2019ply,Goodsell:2019zfs,Domingo:2020wiy} for recent fixed-order calculations).
\item In the effective field theory (EFT) approach, all heavy particles are integrated out and the Higgs mass is then calculated in the low-energy EFT (see~\cite{Hahn:2013ria,Draper:2013oza,Bagnaschi:2014rsa,Vega:2015fna,Lee:2015uza,Bagnaschi:2017xid,Bahl:2018jom,Harlander:2018yhj,Bagnaschi:2019esc,Murphy:2019qpm,Bahl:2019wzx,Bahl:2020tuq,Bahl:2020jaq} for recent EFT calculations).
\item While the FO approach is precise for low but not for high SUSY-breaking mass scales (leading to light or heavy superpartner particles, respectively), it is vice versa for the EFT approach. To obtain a precise prediction also for intermediary scale, hybrid approaches combining both approaches have been developed (see~\cite{Hahn:2013ria,Bahl:2016brp,Athron:2016fuq,Staub:2017jnp,Bahl:2017aev,Athron:2017fvs,Bahl:2018jom,Bahl:2018ykj,Harlander:2019dge,Bahl:2019hmm,Bahl:2019wzx,Kwasnitza:2020wli,Bahl:2020tuq}).
\end{itemize}

In this paper, we will follow the hybrid approach. We combine the results of the recent EFT calculation assuming a THDM as low-energy theory and allowing for non-vanishing phases of the complex parameters~\cite{Pilaftsis:1999qt,Carena:2000yi,Carena:2001fw,Carena:2015uoe,Murphy:2019qpm}  with the fixed-order calculation provided by \FH \cite{Heinemeyer:1998yj,Heinemeyer:1998np,Degrassi:2001yf, Brignole:2001jy,Brignole:2002bz,Degrassi:2002fi,Dedes:2003km, Heinemeyer:2004xw,Frank:2006yh,Heinemeyer:2007aq,Hahn:2009zz, Hollik:2014wea,Hollik:2014bua,Hollik:2015ema,Hahn:2015gaa}. The non-vanishing phases allow for \CP-violating effects. We also include some further improvements of the EFT calculation. On the one hand, this will increase the validity range of the prediction of the Higgs-boson mass as implemented in \FH. On the other hand, it allows for a better prediction of the mixing of the Higgs bosons, which is important for the calculation of partial decay widths and production cross sections, since the THDM allows for a consistent description of the mixing of all the Higgs bosons. The combination can improve the predictions that have been exploited when studying not only a \CP-odd admixture to the lightest Higgs boson but also the \CP-violating mixing between the heavy Higgs bosons and corresponding changes of exclusion bounds~\cite{Bahl:2018zmf}.

This paper is organized as follows: In \Sec{sec:02_EFTimprovements} we discuss the implemented EFT calculation. The combination with the fixed-order calculation is then explained in \Sec{sec:03_FOcomb}. We present numerical results in \Sec{sec:04_results} and conclusions in \Sec{sec:05_conclusions}. Explicit expressions for threshold corrections are given in \App{app:06_thresholds}.

%% file: 02_EFTimprovements.tex
Our EFT calculation closely follows the one published in~\cite{Murphy:2019qpm}. At the scale \msusy, which we choose to be the geometric mean of the stop SUSY soft-breaking mass parameters, we integrate out all sfermions as well as the gauginos and Higgsinos (i.e., we do not take into account separate electroweakino (EWino) and gluino thresholds as in~\cite{Bahl:2018jom} for real parameters). The resulting EFT is the complex THDM (cTHDM) with the Higgs potential,
\begin{align}\label{HiggsPotential}
V_{\text{THDM}}(\Phi_1,\Phi_2) =& m_{11}^2\,\Pdd + m_{22}^2\,\Puu - \left(m_{12}^2 \Pdu + \rm{h.c.}\right) \nonumber\\
& + \frac{1}{2}\lambda_1 (\Pdd)^2 + \frac{1}{2}\lambda_2 (\Puu)^2  + \lambda_3 (\Pdd)(\Puu) + \lambda_4 (\Pdu)(\Pud)\nonumber\\
& + \left(\frac{1}{2}\lambda_5 (\Pdu)^2 + \lambda_6 (\Pdd)(\Pdu) + \lambda_7 (\Puu)(\Pdu) + \rm{h.c.}\right),
\end{align}
where $\Phi_{1,2}$ are the two Higgs doublets. The Higgs mixing parameter $m_{12}^2$ as well as the quartic couplings $\lambda_5$, $\lambda_6$ and $\lambda_7$ are potentially complex parameters while the mass parameters $m_{11}^2$,  $m_{22}^2$ and the quartic couplings $\lambda_1$ to $\lambda_4$ are real.

The Yukawa part of the Lagrangian reads
\begin{align}
\mathcal{L}_{\rm{Yuk}} = - h_t \bar t_R \left(-i \Phi_2^T\sigma_2\right)Q_L - h_t' \bar t_R \left(-i \Phi_1^T\sigma_2\right)Q_L + \rm{h.c.}
\end{align}
with $Q_L$ being the doublet of the left-handed quarks and $t_R$ the right-handed quark singlet. Both top-Yukawa couplings, $h_t$ and $h_t'$, can be complex. In this case, a real top-quark mass can be obtained by a redefinition of the right-handed top-quark field. We neglect the bottom-Yukawa couplings (which were taken into account in~\cite{Murphy:2019qpm}). In the fixed-order calculation of \FH and thereby also in in the fixed-order part of our combined hybrid calculation, the bottom-Yukawa coupling of the MSSM is, however, fully taken into account at the one- and two-loop level. These two-loop fixed-order corrections are evaluated in the gaugeless limit and interpolated for non-vanishing phases (see~\cite{Bahl:2018qog} for details). As investigated in detail in~\cite{Bahl:2020tuq} for the case of the SM as EFT, the resummation of corrections proportional to the bottom-Yukawa coupling beyond the order of the fixed-order calculation becomes relevant only for $\tan\beta\gtrsim 25$ and negative $\mu$.\footnote{We expect very similar results for the THDM as EFT, since a large hierarchy between $M_{H^\pm}$ and \msusy is tightly constrained by collider searches for heavy Higgses decaying to tau leptons (see e.g.\ \cite{Bahl:2018zmf}) for large $\tan\beta$.}

After fixing the cTHDM parameters at \msusy by matching to the full MSSM, we evolve the couplings down to the scale of the non-SM Higgs bosons, which we take to be the charged Higgs mass \MHp. For this evolution, we use the renormalization group equations (RGEs) derived in~\cite{Murphy:2019qpm} following~\cite{Vaughn:1983,Vaughn:1984,Vaughn:1985,Luo:2002ti,Schienbein:2018fsw,Sperling:2013xqa,Sperling:2013eva} and compared to the results of~\cite{Oredsson:2018yho}. At the scale \MHp, we integrate out the heavy Higgs bosons and recover the SM as EFT.

We improve the calculation presented in~\cite{Murphy:2019qpm} by the following aspects: We take into account
\begin{itemize}
\item full one-loop threshold corrections between the SM and the complex THDM (i.e., for the SM Higgs self-coupling and the top-Yukawa coupling),
\item the purely electroweak contributions to the threshold corrections of the THDM Higgs self-couplings at the SUSY scale (i.e., contributions from electroweakinos and the change from \DR, used in the MSSM, to \MS, used in the THDM),
\item \order{\alt\als} contributions to the threshold corrections of the THDM Higgs self-couplings at the SUSY scale,
\item electroweak contributions to the threshold corrections of the THDM top-Yukawa couplings at the SUSY scale.
\end{itemize}
Explicit expressions for these extended threshold corrections are listed in \App{app:06_thresholds}.

Moreover, for obtaining pure EFT results, we improve the extraction of the physical SM-like Higgs mass at the electroweak scale. In~\cite{Murphy:2019qpm}, only the leading \order{\alt} SM corrections were included. We take into account full one- as well as two-loop corrections~\cite{Buttazzo:2013uya} for the SM-like Higgs boson.

As in~\cite{Bahl:2018jom,Murphy:2019qpm}, we calculate all threshold corrections in the limit of degenerate sfermion and electroweakino masses. More explicitly, we assume all sfermions and the gluino to have masses equal to \msusy. Moreover, we set $M_1 = M_2 = \mu$ (where $M_1$ and $M_2$ are the bino and wino soft SUSY-breaking mass and $\mu$ is the Higgsino mass parameter). When the EFT calculation is combined with the fixed-order calculation in the hybrid approach, the effect of non-degenerate masses is, however, fully captured at the order of the fixed-order calculation.

%% file: 03_FOcomb.tex
In this Section, we discuss how the EFT calculation, presented in \Sec{sec:02_EFTimprovements}, is combined with the existing fixed-order calculation implemented in \FH. This fixed-order calculation incorporates full one-loop as well as \order{\alt\als,\alb\als,\alt^2,\alt\alb,\alb^2} two-loop corrections~\cite{Heinemeyer:1998yj,Heinemeyer:1998np,Degrassi:2001yf, Brignole:2001jy,Brignole:2002bz,Degrassi:2002fi,Dedes:2003km, Heinemeyer:2004xw,Frank:2006yh,Heinemeyer:2007aq,Hahn:2009zz, Hollik:2014wea,Hollik:2014bua,Hollik:2015ema,Hahn:2015gaa} ($\alpha_{b,t} = h_{b,t}^2/(4\pi)$ with $h_{b,t}$ being the bottom and top-Yukawa couplings in the MSSM; $\alpha_s = g_3^2/(4\pi)$ with $g_3$ being the strong gauge coupling). The phases of complex parameters are fully taken into account at the one-loop level as well as at \order{\alt\als,\alt^2}. The remaining two-loop corrections are interpolated in the case of non-vanishing phases.

\medskip

For the combination of the EFT calculation with the fixed-order calculation, we largely follow the procedure outlined in~\cite{Bahl:2018jom}: We first redefine the Higgs fields of the fixed-order calculation in order to match the normalization of the Higgs fields in the EFT. The prescription of~\cite{Bahl:2018jom}, valid only for the case of real input parameters, was extended to the case of complex input parameters in~\cite{Bahl:2018ykj}. As a second step, we add the individual results of the EFT and the fixed-order calculation for each specific element of the Higgs two-point-vertex-function matrix, $\Gamma_{hHA}$. Additional subtraction terms ensure that terms contained in the EFT as well as the fixed-order calculation are not double-counted,\footnote{We neglect mixing with the Goldstone bosons, which is a subleading two-loop correction.}
\begin{align}
\label{eq:propmatrix}
&\widehat\Gamma_{hHA}(p^2)=  i\left[p^2 \mathbf{1} -
\begin{pmatrix}
m_h^2 & 0     & 0\\
0     & m_H^2 & 0     \\
0     & 0     & m_A^2
\end{pmatrix} +
\begin{pmatrix}
\hat\Sigma_{hh}^{\text{hybrid}}(p^2) &
\hat\Sigma_{hH}^{\text{hybrid}}(p^2) &
\hat\Sigma_{hA}^{\text{hybrid}}(p^2) \\
\hat\Sigma_{hH}^{\text{hybrid}}(p^2) &
\hat\Sigma_{HH}^{\text{hybrid}}(p^2) &
\hat\Sigma_{HA}^{\text{hybrid}}(p^2) \\
\hat\Sigma_{hA}^{\text{hybrid}}(p^2) &
\hat\Sigma_{HA}^{\text{hybrid}}(p^2) &
\hat\Sigma_{AA}^{\text{hybrid}}(p^2)
\end{pmatrix}\right],
\end{align}
with
\begin{align}
\hat\Sigma_{ij}^{\text{hybrid}}(p^2) = \hat\Sigma_{ij}^{\text{FO}}(p^2) + \Delta_{ij}^\text{EFT} - \Delta_{ij}^\text{sub},
\end{align}
where the $\hat\Sigma^{\text{FO}}_{ij}$'s are the fixed-order self-energies, the $\Delta_{ij}^\text{EFT}$'s are the EFT results, i.e. the elements of the corresponding THDM mass matrix, the $\Delta_{ij}^\text{sub}$'s are the subtraction terms (including tree-level and higher-order terms), and $m_h$, $m_H$, $m_A$ are the tree-level masses of the light CP-even, the heavy CP-even, and the CP-odd Higgs boson, respectively. The pole masses $M_{h_1}$, $M_{h_2}$, $M_{h_3}$ with $M_{h_1} \leq M_{h_2} \leq M_{h_3}$ are then determined by finding the poles of this Higgs two-point-vertex-function matrix. In~\cite{Bahl:2018jom}, only the logarithms in the \CP-even $h,H$-submatrix have been resummed in the hybrid approach.

\medskip

In comparison to the real THDM (rTHDM) hybrid calculation presented in~\cite{Bahl:2018jom}, we change the treatment of the \order{\alt^2} fixed-order corrections. The reason for this change is the following: These \order{\alt^2} fixed-order corrections are by default parameterized in terms of the SM \MS top-quark mass $\overline{m}_t (M_t)$ evaluated at the on-shell top-quark mass $M_t$. They include non-logarithmic SUSY contributions which are constant if \msusy is varied (keeping $\xt = X_t/\msusy$ fixed where $X_t$ is the top squark mixing parameter). For the case of the THDM as EFT, these fixed-order corrections have no EFT counterpart, since the \order{\alt^2} threshold corrections between the SM and the THDM, which have been unknown until recently (see~\cite{Bahl:2020jaq}), are not implemented in our current setup.\footnote{In the scenarios discussed in \Sec{sec:04_results}, the numerical impact of the \order{\alt^2} threshold corrections is expected to be small (see~\cite{Bahl:2020jaq}). The treatment of the fixed-order correction as discussed here is numerically more relevant.}

Consequently, in~\cite{Bahl:2018jom} no corresponding non-logarithmic subtraction term was included. This resulted into a discrepancy between the hybrid results using the THDM and the SM as EFT for high \msusy ($\sim 100\tev$) and $M_A = \msusy$, where one would expect a good agreement of the two calculations. For such a high \msusy, the \order{\alt^2} threshold correction between the SM and the MSSM are actually negligible, since the top-Yukawa coupling shrinks with rising \msusy (also the \order{\alt^2} matching conditions between the THDM and the MSSM will show the same behaviour). For the combined THDM hybrid result of~\cite{Bahl:2018jom}, in contrast, the non-logarithmic \order{\alt^2} fixed-order terms, employing the SM \MS top-quark mass $\overline{m}_t (M_t)$ instead of the top-Yukawa coupling at the scale \msusy, non-negligibly contribute to the final result. This shows clearly that for very high \msusy, the \order{\alt^2} fixed-order terms should not be included if the EFT counterpart (the \order{\alt^2} threshold correction and the corresponding subtraction term) is not available. For low \msusy, the inclusion of the fixed-order \order{\alt^2}, however, can improve the result. Since, in this paper, we are mainly interested in the low $\tan\beta$ region,\footnote{In the high $\tan\beta$ region (i.e.\ $\tan\beta\gtrsim 10$) a large hierarchy between the SUSY scale and the scale of the heavy Higgs bosons is excluded due to an interplay of the SM-like Higgs mass constraint and tight constraints from heavy Higgs searches (see e.g.~\cite{Bahl:2018zmf,Bahl:2019ago}). Since in addition Higgs mixing effects are suppressed for large $\tan\beta$, using the SM as EFT provides a very accurate approximation of the THDM-EFT result~\cite{Bahl:2018jom}.} where a high \msusy is needed to obtain $M_h \simeq 125\gev$, we chose to subtract the SUSY \order{\alt^2} contribution to the $hh$ Higgs self-energy.

\medskip

An additional step has to be considered for the combination of the fixed-order and the EFT result if the input parameters of the fixed-order calculation and the EFT calculation are not renormalized in the same scheme. In this context, the stop mixing parameter, $X_t$, is especially relevant. While $X_t$ can be renormalized either in the \DR or the OS scheme in case of the fixed-order calculation, in the EFT calculation $X_t$ is renormalized in the \DR scheme. In case of an OS input parameter, we need to convert $X_t$ from the OS to the \DR scheme. As argued in \cite{Bahl:2016brp,Bahl:2017aev}, a conversion of  $X_t$ taking into account only large logarithmic terms is sufficient. For the present study, we generalize the formula given in~\cite{Bahl:2018jom}, which is valid in case of a large hierarchy between the non-SM Higgs scale and the SUSY scale, to the case of complex input parameters (see also~\cite{Bahl:2020tuq}),
\begin{align}
X_t^\DR(\msusy) &= X_t^\OS\left\{1 + \left[\frac{\als}{\pi} - \frac{3\alt}{16\pi}\left(1 - |\hat X_t|^2\right)\right]\ln\frac{\msusy^2}{M_t^2} \right.\nonumber\\
&\left.\hspace{1.93cm} + \frac{3}{16\pi}\frac{\alt}{\tbb}\left(1 - |\hat Y_t|^2\right)\ln\frac{\msusy^2}{M_{H^\pm}^2}\right\},
\end{align}
where $\hat X_t = X_t/\msusy$ and $\hat Y_t = \hat X_t + 2\hat\mu^*/s_{2\beta}$ (with $\hat\mu = \mu/\msusy$ and $\mu$ being the Higgsino mass parameter) and $t_\gamma \equiv \tan\gamma$ and $s_\gamma\equiv\sin\gamma$ are used for abbreviation.

%% file: 04_results.tex
In this Section, we numerially investigate the improved EFT calculation as well as the combined hybrid calculation. For most of the numerical results, we consider a simple two-scale scenario. All sfermion SUSY soft-breaking mass parameters $M_{\tilde{f}}$ as well as the gaugino and Higgsino mass parameters, $M_1$, $M_2$, $M_3$, $\mu$, are set equal to \msusy. As second scale, we allow the mass of the charged Higgs boson \MHp to be different from \msusy. We set all SUSY-soft trilinear couplings except of $A_t$ to zero. In summary,
\begin{align}\nonumber
M_{\tilde{f}} &= \msusy, \quad M_1 = M_2 = M_3 = \msusy, \quad \mu = \msusy, \\
&A_{f\neq t} = 0, \quad A_t = X_t + \mu^*/\tbe,
\end{align}
where $X_t$ is the top squark mixing parameter.

If not stated otherwise, all parameters are fixed in the \DR scheme at the scale \msusy (apart from \MHp, which we fix in the \MS scheme at the scale \MHp and in case of a pure EFT calculation and in the OS scheme in case of a hybrid calculation, and $\tan \beta$, which is always fixed in the \MS scheme at the scale \MHp) and all phases are set to zero.

\medskip

\afterpage{
\begin{figure}
\begin{minipage}{.48\textwidth}\centering
\includegraphics[width=\textwidth]{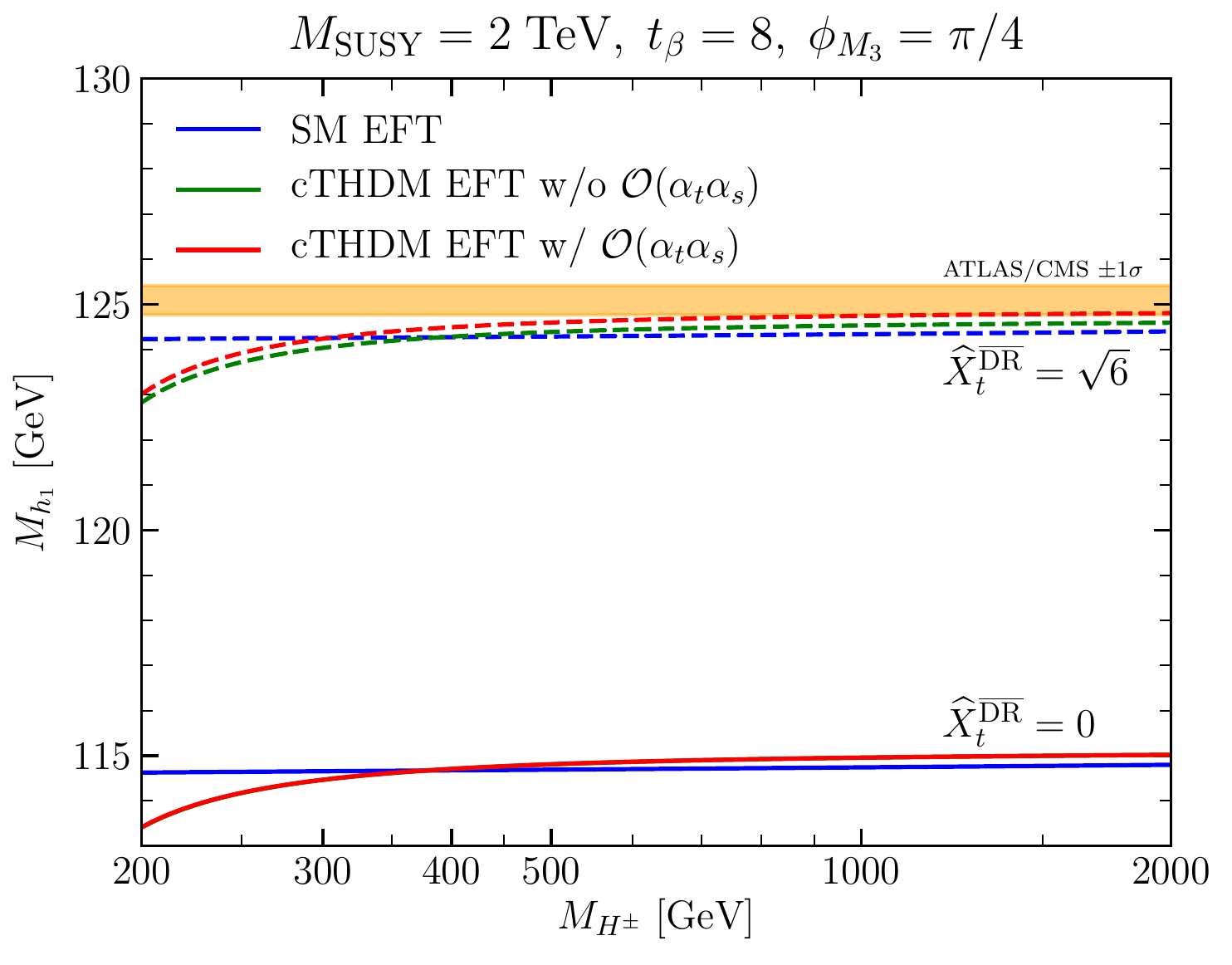}
\end{minipage}
\begin{minipage}{.48\textwidth}\centering
\includegraphics[width=\textwidth]{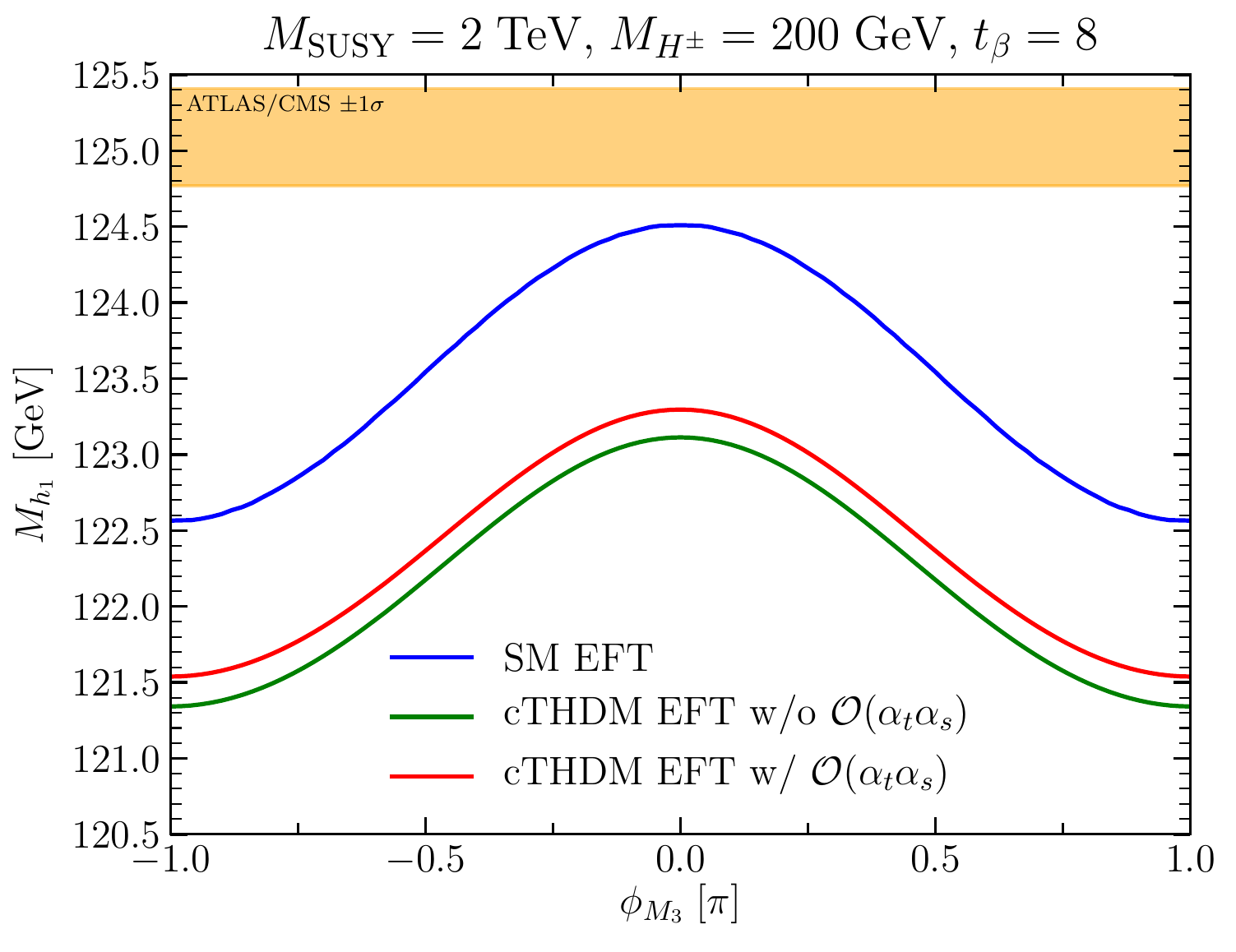}
\end{minipage}
\caption{\textit{Left:} \Mh in dependence of \MHp for $\xt^\DR = 0$ (solid) and $\xt^\DR = \sqrt{6}$ (dashed). The result obtained using a SM-EFT calculation (blue) are compared to results using cTHDM-EFT calculations not including the \order{\alt\als} threshold corrections (green) and including the \order{\alt\als} threshold corrections (red). The yellow band displays the Higgs mass as measured by ATLAS and CMS within one standard deviation of experimental accuracy.\protect\footnotemark \textit{Right:} Same as left plot but \Mh is shown in dependence of \pMiii for $\xt^\DR = \sqrt{6}$.}
\label{fig:EFT}
\end{figure}
\footnotetext{We do not include a band for the theoretical uncertainty, since a reliable uncertainty estimate requires further investigations (for the case of the SM as EFT see, for example, Ref.~\cite{Bahl:2019hmm}). Moreover, the main purpose of the shown plots is to investigate the dependencies of $M_h$ on the model parameters and not to employ $M_h$ to constrain these parameters.}
}

First, we assess the numerical size of taking into account the \order{\alt\als} threshold corrections in the left plot of \Fig{fig:EFT}. It shows the mass of the lightest Higgs boson \Mh in dependence on \MHp. The mass \Mh is calculated in the pure EFT approach using three different EFTs: SM EFT (blue), cTHDM EFT without \order{\alt\als} threshold corrections (green) and cTHDM EFT with \order{\alt\als} threshold corrections (red). The SM-EFT result is based on the calculation presented in~\cite{Bahl:2020tuq} including the full phase dependence in the threshold corrections between the SM and the MSSM.

We choose $\msusy=2\tev$, $\tbe=8$ and $\pMiii = \pi/4$. For smaller $\tbe$, higher \msusy values would be required to obtain $\Mh\sim 125\gev$. Since the value of the strong gauge coupling, entering the calculation evaluated at \msusy, shrinks with rising \msusy, the numerical impact of the \order{\alt\als} threshold corrections would be even smaller.

For $\xt^\DR = 0$, the numerical effect of the \order{\alt\als} threshold corrections is completely negligible. The THDM-EFT results with and without these corrections lie on top of each other. For $\xt^\DR = \sqrt{6}$, there is a small approximately constant difference between both curves ($\sim 0.2\gev$). The shift induced by the \order{\alt\als} threshold corrections is relatively small in comparison to results obtained in the literature for the case of the SM as EFT (see e.g.\ \cite{Bagnaschi:2014rsa,Draper:2013oza,Bahl:2016brp}). This is explained by the fact that we express the \order{\alt} threshold correction in terms of the MSSM top-Yukawa coupling. As discussed in~\cite{Kwasnitza:2020wli}, this choice absorbs the numerically most significant two-loop terms (i.e., those terms including the highest powers of $\at$ and $\mf$).

The SM-EFT result is in good agreement with the THDM-EFT results for $\MHp \gtrsim 300\gev$. The remaining small difference in this region is explained by the fact that we do not include \order{\alt^2} threshold corrections in the THDM EFT, which have ony been calculated very recently \cite{Bahl:2020jaq}, while the SM EFT includes the \order{\alt^2} threshold correction between the SM and the MSSM. For $\MHp \lesssim 300\gev$, the SM-EFT result lies up to $\sim1\gev$ higher than the THDM-EFT results. This is explained by sizeable Higgs-mixing effects which are taken into account in the THDM-EFT results but not in the SM-EFT results. The numerically effect of the Higgs mixing is stronger for lower $\tbe$ values~\cite{Bahl:2018jom}.

It should be noted that low values of $\MHp$ below $\MHp \approx 500$ GeV are experimentally disfavoured (see for example the phenomenological studies in~\cite{Arbey:2017gmh,Bahl:2018zmf,Bahl:2019ago}). Therefore, our numerical investigation in the left plot of \Fig{fig:EFT} can be regarded as a conservative estimate of the expected maximal size of the effect.

To assess the sensitivity of \Mh on the phases, we compare the different EFT predictions in the right plot of \Fig{fig:EFT} in dependence on \pMiii for $\MHp = 200 \gev$ and $\xt^\DR = \sqrt{6}$. The line coloring is the same as in the left plot of \Fig{fig:EFT}.

All three curves show a quite similar dependence on $\pMiii$. The SM curve is shifted upwards with respect to the cTHDM curves by $\sim 1\gev$, since Higgs mixing effects are not included. Including the \order{\alt\als} threshold corrections in the cTHDM calculation leads to an upwards shift of $\sim 0.2\gev$ but does not introduce any sizeable new phase dependence. For smaller values of $|\xt^\DR|$, the phase dependence is similar but of smaller magnitude.

\medskip

\begin{figure}
\begin{minipage}{.48\textwidth}\centering
\includegraphics[width=\textwidth]{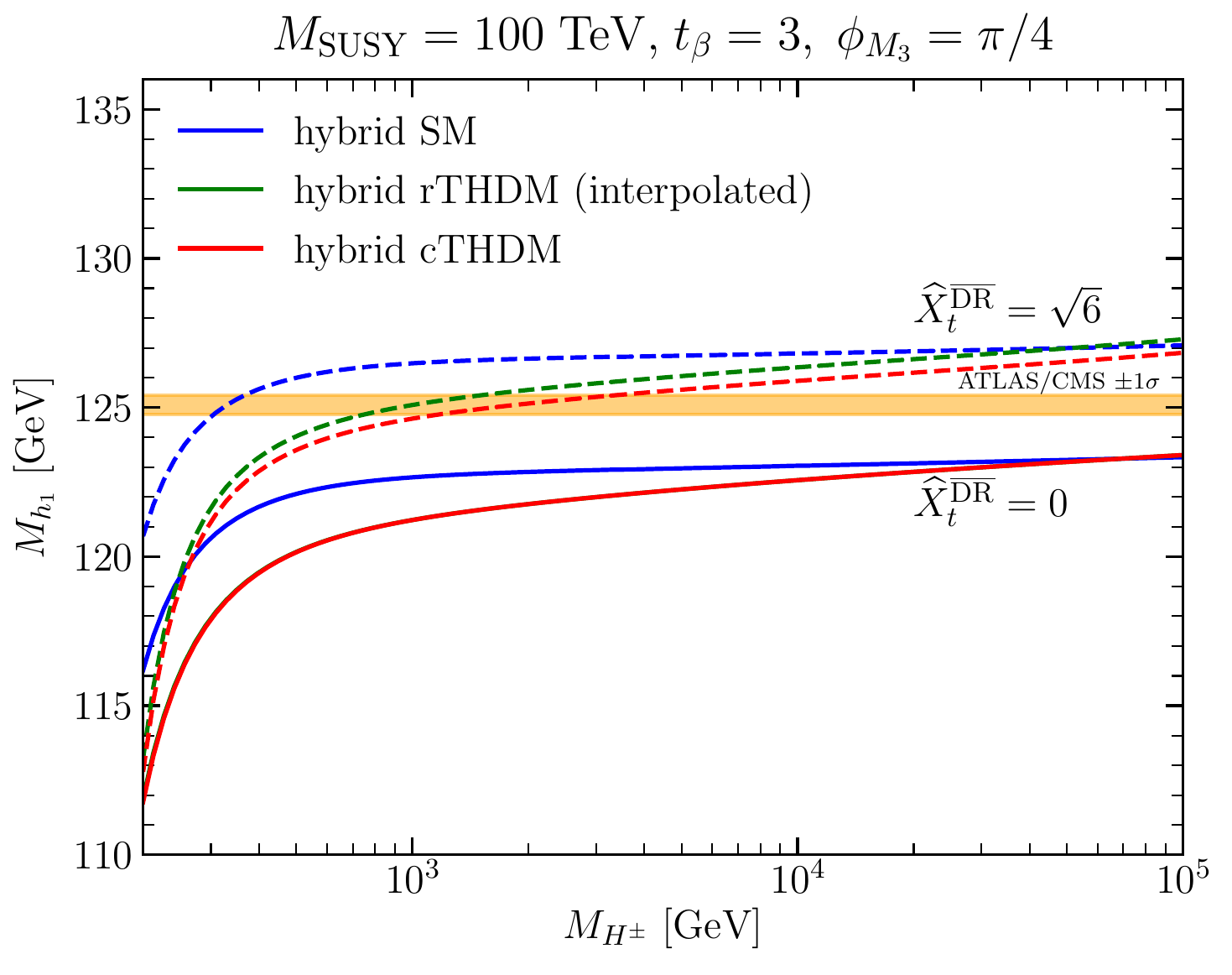}
\end{minipage}
\begin{minipage}{.48\textwidth}\centering
\includegraphics[width=\textwidth]{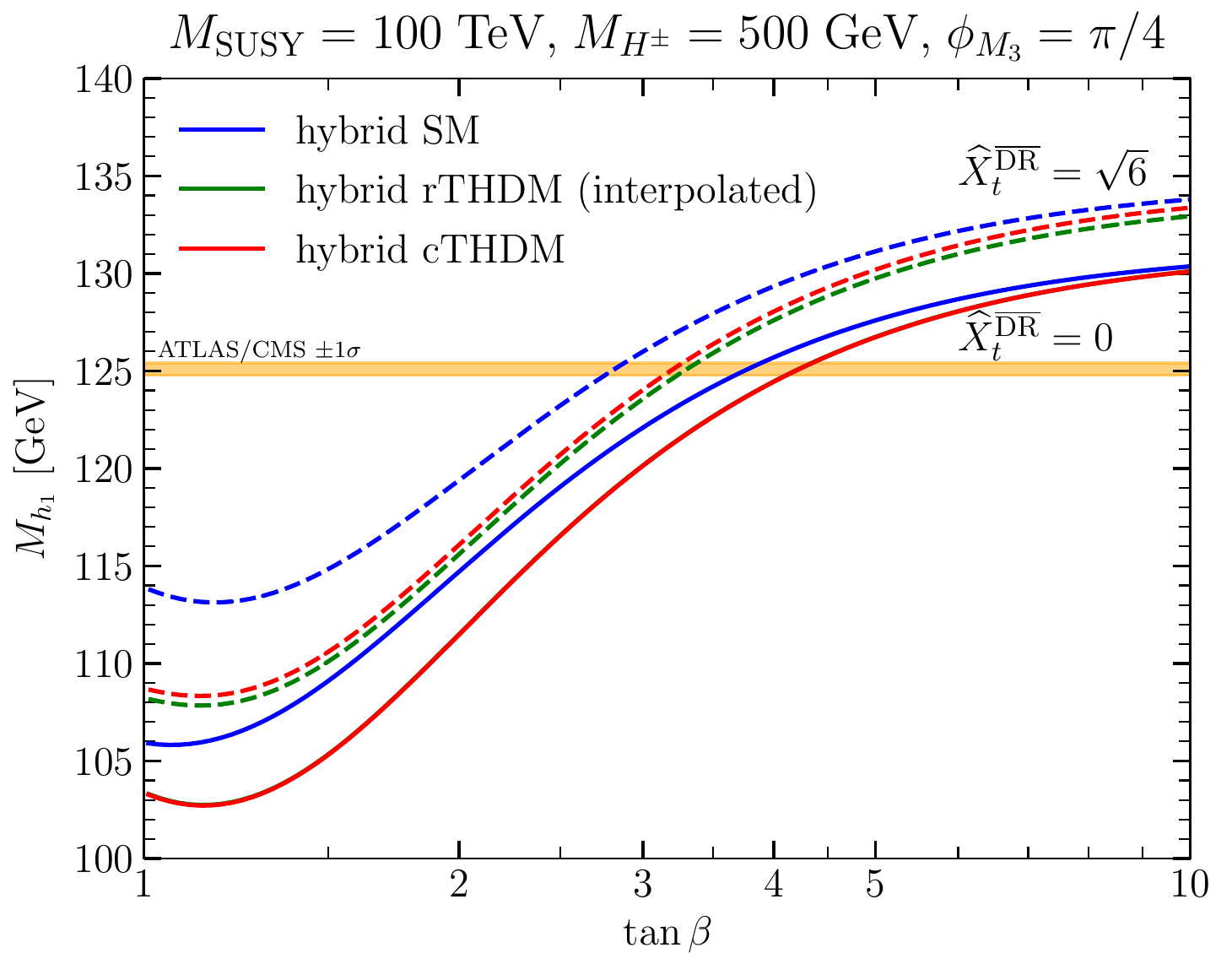}
\end{minipage}
\caption{\textit{Left:} \Mh in dependence of \MHp for $\xt^\DR = 0$ (solid) and $\xt^\DR = \sqrt{6}$ (dashed). The result obtained using the SM hybrid calculation (blue) are compared to results using the interpolated hybrid THDM calculation (green) and the hybrid cTHDM calculation (red). The yellow band displays the Higgs mass as measured by ATLAS and CMS within one standard deviation of experimental accuracy. \textit{Right:} Same as left plot but \Mh is shown in dependence of $\tbe$.}
\label{fig:hybrid_MHp_tb_var}
\end{figure}

In \Fig{fig:hybrid_MHp_tb_var}, we compare the results of hybrid calculations using the SM (blue), the real THDM interpolated in the case of complex input parameters (green)\footnote{This calculation corresponds to the calculation presented in~\cite{Bahl:2018jom} with the modifications detailed in \Sec{sec:03_FOcomb}. The EFT as well as the two-loop fixed-order corrections proportional to the bottom-Yukawa coupling are interpolated for non-vanishing phases.} and the cTHDM (red) as EFTs. For both plots, we set $\msusy = 100\tev$ and $\pMiii = \pi/4$. For the solid lines, $\xt^\DR = 0$ is chosen, while for the dashed lines $\xt^\DR = \sqrt{6}$ is set.

In the left plot, we show \Mh in dependence on \MHp for $\tbe = 3$. For $\xt^\DR = 0$, the THDM results agree very well (the red curve lies on top of the green curve), since the dependence on $\pMiii$ is very small. For $\MHp \sim \msusy$, also the SM result agrees very well with the THDM results. In comparison to \Fig{fig:EFT}, the \order{\alt^2} threshold correction is negligible since the top-Yukawa coupling entering the threshold corrections is very small for $\msusy = 100 \tev$. For $\MHp \ll \msusy$ there is a sizeable difference between the SM and the THDM results again caused by Higgs-mixing effects which become increasingly important if \MHp is lowered. Note that via the combination of the EFT and the fixed-order calculation in the hybrid approach, Higgs-mixing effects are included at the one- and two-loop level also for the result using the SM as EFT. Larger differences between using the SM as EFT and the THDM as EFT would be found in the pure EFT approach.

The general behaviour is similar for $\xt^\DR = \sqrt{6}$. The gluino phase, \pMiii, has, however, a larger numerical impact. This results in an approximately constant shift ($\sim 0.5\gev$) of the rTHDM curve (using interpolation for handling the complex input parameters) with respect to the cTHDM curve.

The right plot of \Fig{fig:hybrid_MHp_tb_var} is analogeous to the left plot but we show \Mh in dependence of $\tbe$ for $\MHp = 500\gev$. The behaviour of the different curves is similar as in the left plot. All results agree very well for $\tbe \sim 10$. For lower $\tbe$, Higgs-mixing effects become relevant leading to a difference between the SM and the THDM results of $\sim 2\gev$ for $\tan\beta \sim 3$. The non-zero gluino phase is only relevant for $\xt^\DR = \sqrt{6}$ leading to a shift of $\sim 0.4\gev$ between the interpolated rTHDM result and the cTHDM results. For $\xt^\DR = 0$, $|\at| \le 1$ since $|\mf| = 1$ is chosen. Since in this case, the dependence on $\pMiii$ is not enhanced by powers of $|\at|$ or $|\mf|$ (see explicit expressions in \App{app:06_thresholds}) the red and green solid lines lie on top of each other.

\medskip

\begin{figure}
\begin{minipage}{.48\textwidth}\centering
\includegraphics[width=\textwidth]{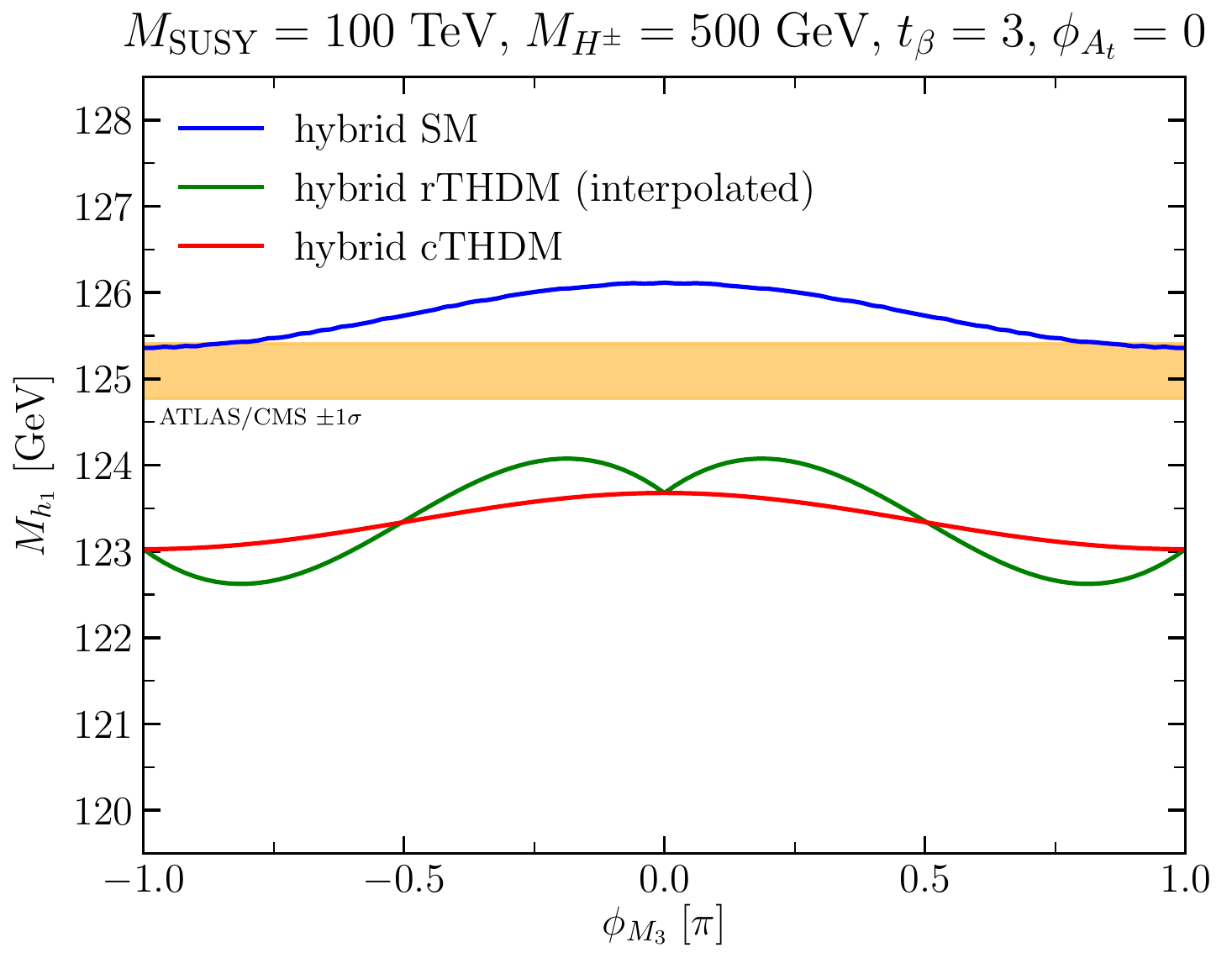}
\end{minipage}
\begin{minipage}{.48\textwidth}\centering
\includegraphics[width=\textwidth]{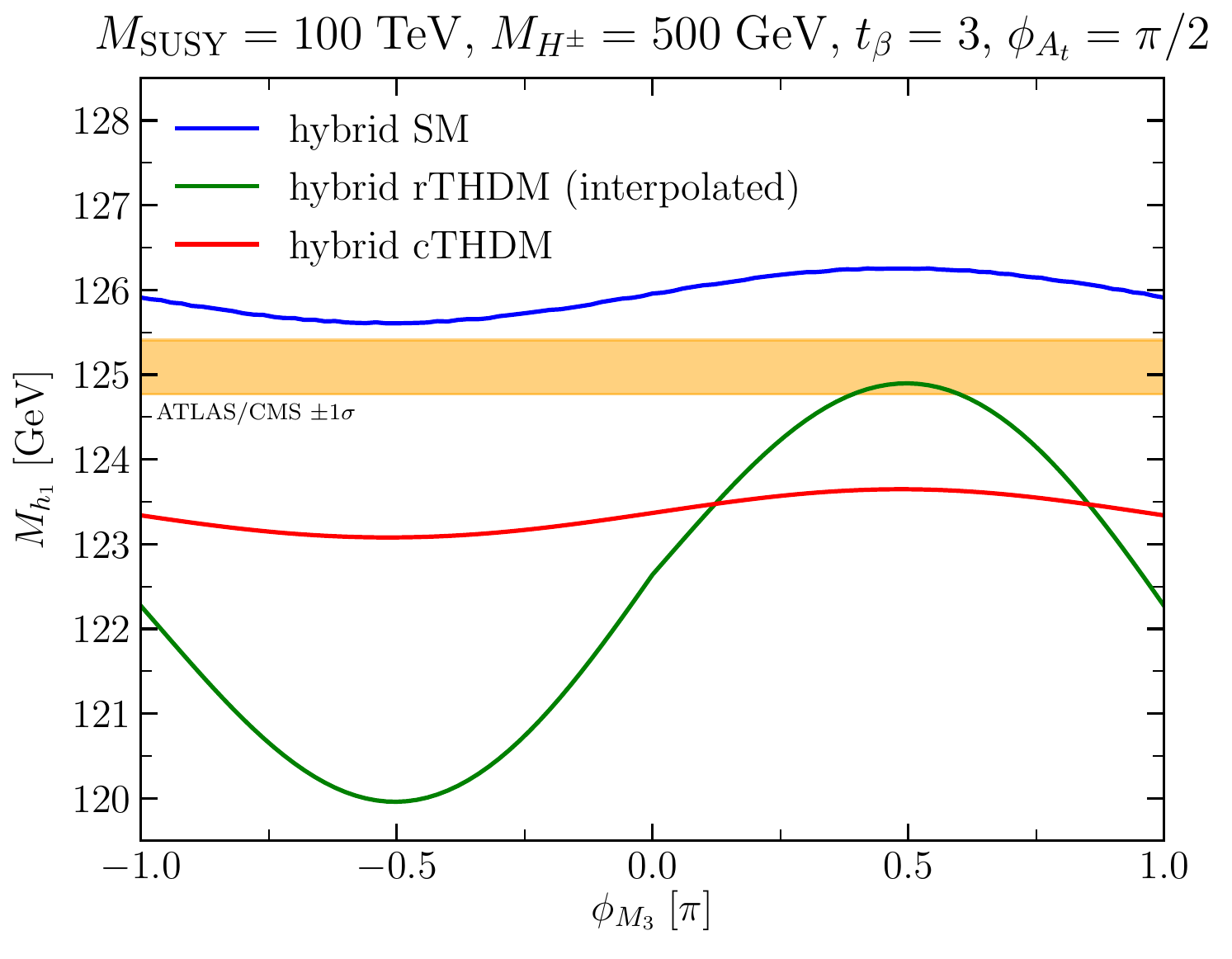}
\end{minipage}
\caption{\textit{Left:} \Mh in dependence of \pMiii for $\pAt = 0$ and $\xt = \sqrt{6}$. The result obtained using a the SM hybrid calculation (blue) are compared to results using the interpolated hybrid THDM calculation (green) and the hybrid cTHDM calculation (red). The yellow band displays the Higgs mass as measured by ATLAS and CMS within one standard deviation of experimental accuracy. \textit{Right:} Same as left plot but $\pAt = \pi/2$.}
\label{fig:hybrid_pMiii_var}
\end{figure}

We further investigate the dependence on the the phases in \Fig{fig:hybrid_pMiii_var}. The same hybrid calculations as in \Fig{fig:hybrid_MHp_tb_var} are shown in dependence on \pMiii for $\msusy = 100 \tev$, $\MHp = 500\gev$, $\tbe = 3$ and $\xt^\DR = \sqrt{6}$.

In the left plot, we set $\pAt = 0$. The SM and the cTHDM curves are approximately parallel to each other. The off-set between both curves is caused by Higgs-mixing effects, which are fully taken into account for the cTHDM curve while for the SM EFT results the Higgs-mixing corrections are only included via the fixed-order corrections. The interpolated rTDHM curve agrees as expected with the cTHDM curve for $\pMiii = 0,\pm\pi$. In between these phase values, the interpolation of the EFT part of the calculation leads to shifts of $\sim 0.5\gev$. The kink at $\pMiii = 0$ originates from the combination of the fully phase dependent fixed-order calculation and the interpolated rTHDM EFT calculation (see also \cite{Bahl:2018qog,Bahl:2020tuq}).

The difference of the interpolated rTHDM and the cTHDM results is enlarged if a second phase is non-zero. In the right plot of \Fig{fig:hybrid_pMiii_var}, we set $\pAt = \pi/2$. While the non-interpolated results show only a very mild dependence on \pMiii as expected by the fact that the strong gauge coupling entering the threshold corrections at \msusy is small for $\msusy=100\tev$. In contrast, the interpolated rTHDM curve shows a strong dependence on \pMiii leading to difference in comparison to the cTHDM curve of up to $3.5\gev$.

For the scenarios of \Fig{fig:hybrid_pMiii_var}, the dependence on $\pAt$ is very similar to the one on $\pMiii$.

\medskip

\begin{figure}
\begin{minipage}{.48\textwidth}\centering
\includegraphics[width=\textwidth]{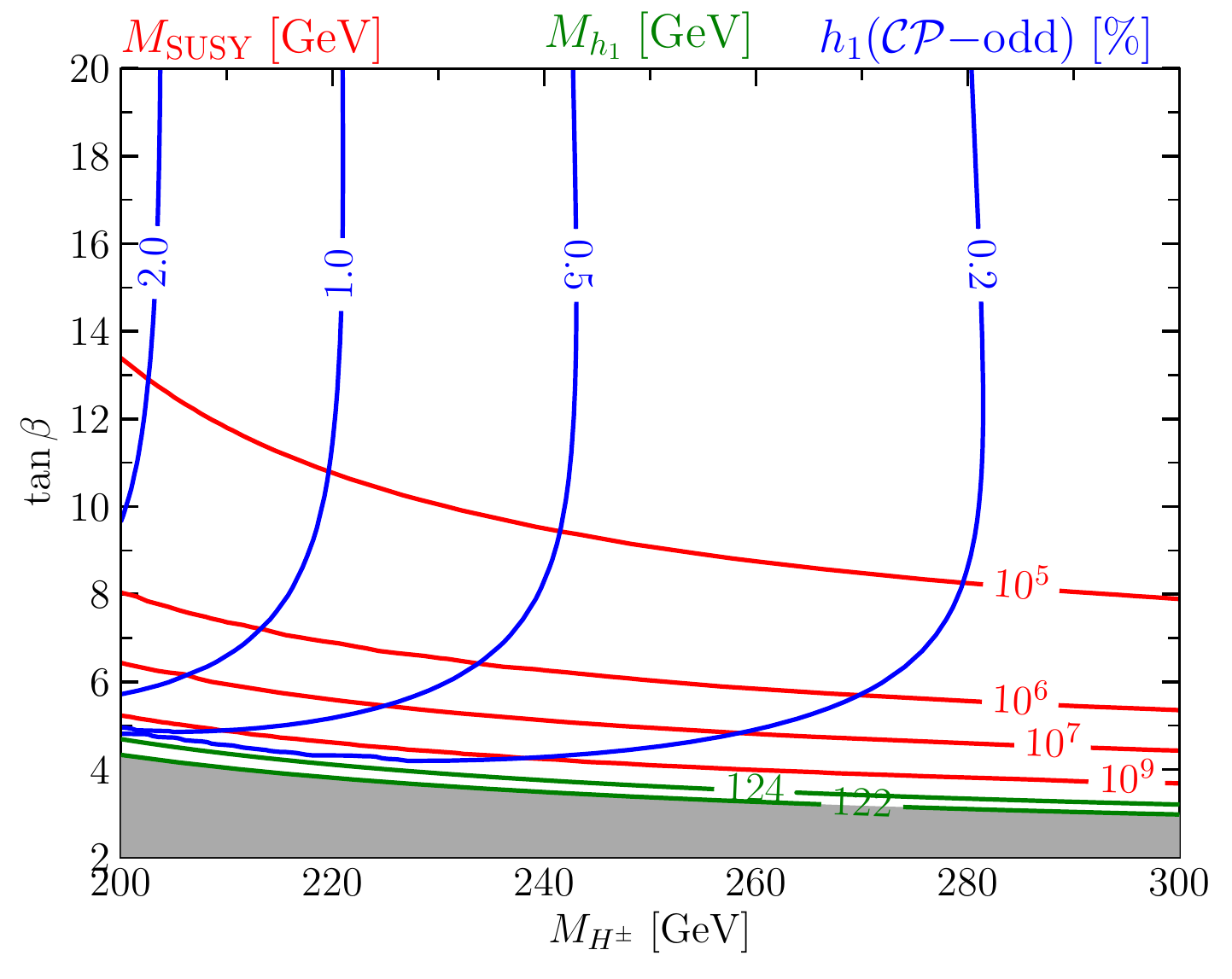}
\end{minipage}
\begin{minipage}{.48\textwidth}\centering
\includegraphics[width=\textwidth]{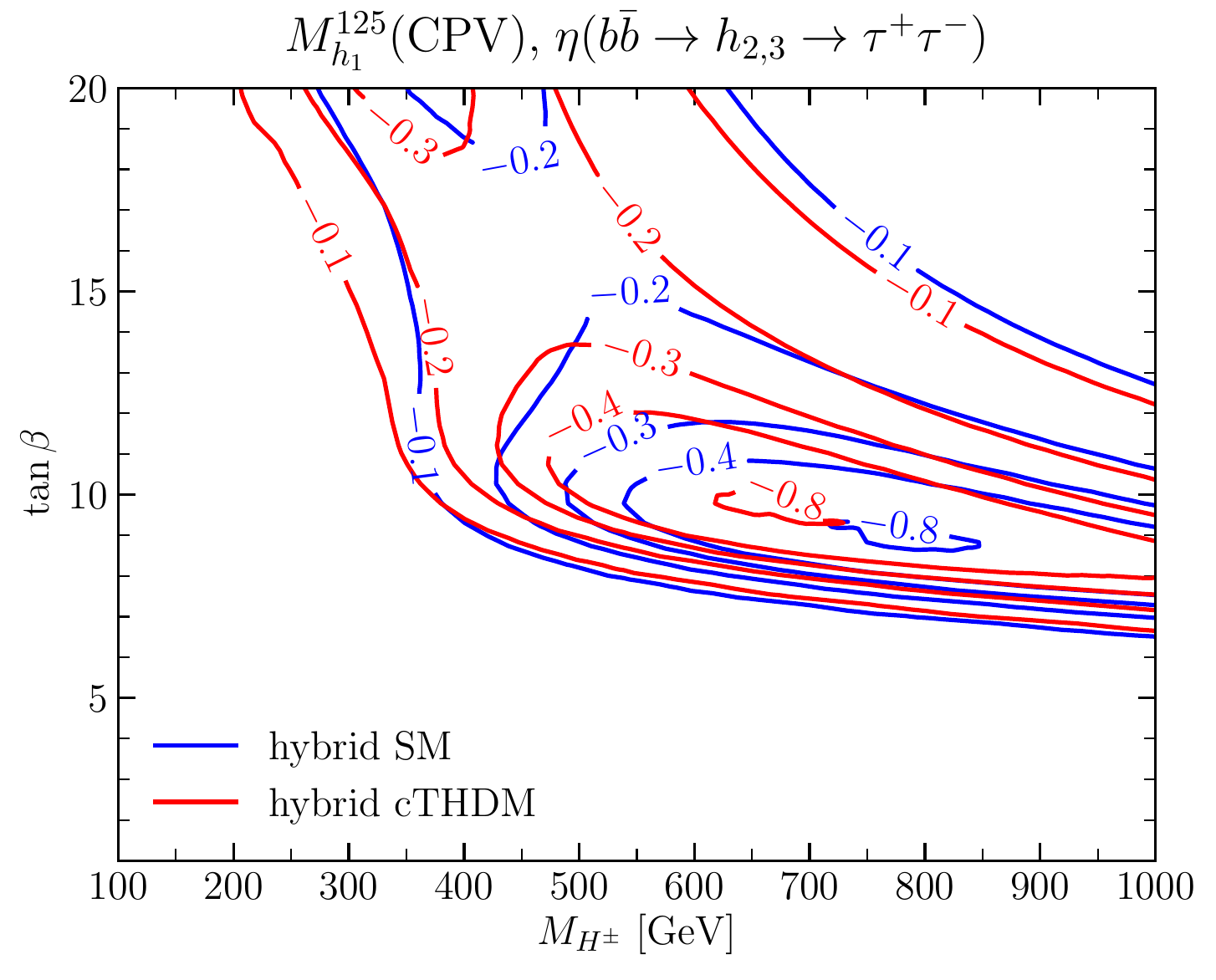}
\end{minipage}
\caption{\textit{Left:} The contour of $\msusy$ (red), of $\Mh$ (green) and, of the size of the \CP-odd component of $h_1$ (blue) is shown in the $\MHp-\tan \beta$ plane for $|\at| = 3$, $|\mf| = 3$, $\pAt = 2\pi/3$ and $\pMiii = \pi/4$. \textit{Right:} The contours of the interference factor $\eta$  in the $\MHp-\tan \beta$ plane are compared applying the SM (blue) or the cTHDM as EFT in the hybrid approach for the $M_h^{125}(\rm{CPV})$ benchmark scenario as defined in~\cite{Bahl:2018zmf}.}
\label{fig:pheno}
\end{figure}

Next, we study phenomenological applications of our cTHDM calculation. In the left plot of \Fig{fig:pheno}, we try to assess the maximally allowed \CP-odd component of the SM-like Higgs bosons. Following~\cite{Murphy:2019qpm}, we evaluate the size of the component by the $(1,3)$ entry of the Higgs mixing matrix squared. We employ a similar scenario as used in~\cite{Murphy:2019qpm}. That means we set $|\at| = 3$, $|\mf| = 3$, $\pAt = 2\pi/3$ and $\pMiii = \pi/4$. For every point in the $(\MHp,\tbe)$ parameter plane we adjust \msusy such that our pure cTHDM EFT calculation yields $\Mh \sim 125\gev$. As upper limit on \msusy, we employ $10^{16}\gev$.\footnote{A similar procedure has been applied in~\cite{Bahl:2019ago}} After fixing \msusy, we calculate the \CP-odd component of the SM-like Higgs boson as outlined in~\cite{Murphy:2019qpm}.

We show three different sets of contours: $\msusy$ contours (red), \Mh contours (green) and the \CP-odd component of the SM-like Higgs (blue). The minimal value of \msusy is reached for high $\tbe$ and \MHp in the upper right corner of the plot. Lowering $\tbe$ and \MHp implies the need to increase \msusy in order to guarantee $\Mh \sim 125\gev$. For $\tbe \sim 3-4.5$, the maximum \msusy value of $10^{16}\gev$ is reached. In the same region, \Mh drops below 125~GeV. The gray-shaded region, in which \Mh is lower than 122 GeV, is excluded by the experimentally measured Higgs mass (assuming a conservatively estimated theoretical uncertainty of $\sim 3\gev$).

The size of the \CP-odd component of the SM-like Higgs is highest for large $\tbe$ and small \MHp. This is seemingly in contrast to the findings in~\cite{Murphy:2019qpm}, where the largest \CP-odd component was found for low $\tbe$ and small \MHp, where Higgs-mixing effects are most important. In~\cite{Murphy:2019qpm}, however, a fixed SUSY scale of 30~TeV was considered. Correspondingly, \Mh dropped below 122 GeV already for $\tbe \lesssim 8-10$. Increasing the SUSY scale in this region leads to smaller threshold corrections, since the stronge gauge as well as the top-Yukawa coupling decrease with rising \msusy. Since these threshold corrections induce the \CP-violation, larger \msusy values directly result into a smaller \CP-odd component of the SM-like Higgs boson.

The considered parameter space is already tightly constraints by experimental searches for heavy Higgs bosons as well as coupling measurements of the Higgs boson found at the LHC~(see~\cite{Bahl:2018zmf,Bahl:2019ago} for a detailed discussion of the experimental constraints in similar scenarios). This implies that if a non-zero \CP-odd component of the discovered Higgs boson is established at the LHC, it would probably be very difficult to explain such an observation within the MSSM.\footnote{In the high $\tan\beta$ region, \CP-violating effects from the sbottom sector can become important. This region is, however, even tighter constrained as the low $\tan\beta$ region (see e.g.~\cite{Bahl:2018zmf}). Another possibility might be scenarios in which the second lightest MSSM Higgs boson is SM-like. But also these scenarios are tightly constraint (see e.g.~\cite{Bahl:2018zmf}). Note, however, that a more sizeable \CP-odd component of the SM-like Higgs boson can be obtained in models beyond the MSSM (see e.g.~\cite{King:2015oxa}).}

It should be noted that these tight constraints come solely from the Higgs sector. Measurements such as the ones of the electric dipole moments will restrict the allowed values of the \CP-violating phases and might lead to even lower values of an experimentally allowed \CP -odd component of the Higgs boson, see for example~\cite{Arbey:2014msa, Berger:2015eba, Abe:2018qlw, Cesarotti:2018huy, Arbey:2019pdb}.

\medskip

As second phenomenological application, we study the $M_h^{125}(\rm{CPV})$ benchmark scenario defined in~\cite{Bahl:2018zmf}. In this scenario the SUSY scale is set to 2~TeV, $X_t^\OS = 2.8\tev$ and $\pAt = 2\pi/15$. The non-zero $A_t$ phase leads to mixing between the heavy Higgs bosons, which are almost mass-degenerate. Consequently, large interference effects occur in processes involving an $s$-channel exchange of these Higgs bosons. Phenomenologically, the production of the heavy Higgs bosons via $b$-associated production and their subsequent decay into a pair of tau leptons is very relevant, since it leads to stringent constraints on the MSSM Higgs sector.

In~\cite{Bahl:2018zmf}, the interference effects in this channel were calculated using \FH version~\texttt{2.14.3}, employing a hybrid calculation with the SM as EFT,\footnote{In \texttt{FeynHiggs-2.14.3} the phase dependence of the threshold corrections calculated in~\cite{Bahl:2020tuq} was not yet implemented. Instead the EFT calculation was interpolated in case of complex input parameters.}  and \texttt{SusHi}~\cite{Harlander:2012pb,Harlander:2016hcx} with its extension \texttt{SusHiMi}~\cite{Liebler:2016ceh}. The interference was quantified factorizing the experimentally measurable coherent cross-section in the form
\begin{align}
\sigma(b\bar b\rightarrow h_{1,2,3}\rightarrow\tau^+\tau^-) = \sum_{a=1}^3\sigma(b\bar b\rightarrow h_a)(1 + \eta_a^{IF})\text{BR}(h_a\rightarrow\tau^+\tau^-)
\end{align}
with
\begin{align}
\eta_2^{IF} = \eta_3^{IF} \equiv \eta(b\bar b\rightarrow h_{2,3}\rightarrow\tau^+\tau^-).
\end{align}
see~\cite{Fuchs:2014ola,Fuchs:2016swt,Fuchs:2017wkq} for a more details.

Here, we repeat the calculation of the interference factor $\eta$ with our new hybrid calculation using the cTHDM as EFT (implemented in \FH and linked to \texttt{SusHi}). Since the release of \FH version~\texttt{2.14.3}, also the SM-EFT calculation has been updated. We show here the results obtained using the SM-EFT calculation presented in~\cite{Bahl:2020tuq} which includes the full phase dependence in the threshold corrections. In the given scenario, the numerical difference to the interpolated result used in~\cite{Bahl:2018zmf} is very small.

In the right plot of \Fig{fig:pheno}, the results using the SM as EFT (blue) and the cTHDM as EFT (red) are compared. As visible in the plot, only small differences between both results are visible. The point of maximal interference is shifted slightly to lower \MHp values if the THDM is used as EFT. This is mainly due to small shifts in the heavy Higgs masses $M_{h_2}$ and $M_{h_3}$. The two-loop fixed-order corrections, included in both results, already contain the leading logarithms as well as the leading phase dependence. Due to the small hierarchy between \MHp and \msusy, the resummation of higher-order logarithms appearing in the \CP-mixing self-energy is negligible. Therefore, the hybrid result using the SM as EFT is a good approximation of the full hybrid result using the cTHDM as EFT. This shows the benefits of the hybrid framework combining fixed-order and EFT calculations.

%% file: 05_conclusions.tex
\cp\ violation in BSM-Higgs sectors is one intriguing possibility to overcome the lack of \cp\ violation in the SM; this lack renders an explanation of the matter--antimatter asymmetry of the Universe difficult. In this work, we discussed precision predictions in the \cp-violating MSSM using the cTHDM as EFT.

We improved an existing cTHDM-EFT calculation of the MSSM Higgs boson masses by calculating so far missing one-loop threshold corrections as well as by including the leading two-loop \order{\alt\als} threshold corrections. We found that the resulting numerical impact on the lightest SM-like Higgs boson mass was small ($\sim 0.2\gev$), since the dominant pieces are already absorbed into the one-loop threshold correction by parameterizing them in terms of MSSM parameters.

As a next step, we combined this improved EFT calculation with the existing fixed-order calculation implemented in the public code \FH. In this way, a precise prediction of the MSSM Higgs boson masses for low, intermediary and high SUSY scales is possible. In comparison to the already implemented rTHDM-EFT calculation, which is interpolated in the case of non-vanishing \cp-violating phases, we found shifts of several GeV if more than one phase is non-zero.

As a further benefit, the implementation into \FH allows precise predictions for the Higgs decay rates and production cross-sections resumming all large logarithmic contributions in the Higgs self-energies. As example application, we investigate how large the \cp-odd component of the SM-like Higgs boson can become in the MSSM. Since the \cp-odd component can not exceed a few percent, a discovery of a \cp-odd component at the LHC, assuming the expected experimental reach, would most likely rule out the MSSM. As second application, we studied the interference effect between the two heavier Higgs bosons in the $M_h^{125}(\rm{CPV})$ Higgs benchmark scenario, which was defined in~\cite{Bahl:2018zmf}. In comparison to using the SM as EFT, the use of the cTHDM as EFT did not significantly shift the interference pattern indicating that the two-loop fixed-order corrections already account for the bulk of the corrections. This demonstrate the advantages of the hybrid approach combining fixed-order and EFT calculations.

The presented calculation will become part of an upcoming \FH version.

%% file: 06_app_thresholds.tex
In this Appendix, we list all threshold corrections taken into account in addition to those presented in~\cite{Murphy:2019qpm}.


\subsection{Matching the SM to the cTHDM}
\label{SMtoTHDM}

The SM Higgs self-coupling is obtained in terms of the $\lambda_i$ of the THDM by
\begin{align}\label{lambdaSMtoTHDM}
\lambda(M_{H^\pm}) =& \lambda_{\text{tree}} + \Delta\lambda_{\Re} + \Delta\lambda_{\Im}
\end{align}
with
\begin{subequations}
\begin{align}
\lambda_{\text{tree}} =& \li \cbe^4 + \lii \sbe^4 + 2 (\liii+\liv+\lvRe)\cbb\sbb + 4\lviRe \cbe^3\sbe+4\lviiRe \cbe\sbe^3, \label{eq:SMtoTHDMtree}\\
\Delta_{\Re}\lambda =& - 3k\Big((\lviRe + \lviiRe)c_{2\beta}+ (\lviRe - \lviiRe)c_{4\beta} \nonumber\\
&\hspace{1.3cm}-\left(\li\cbb - \lii\sbb - (\liii+\liv+\lvRe)c_{2\beta}\right)s_{2\beta}\Big)^2, \\
\Delta_{\Im}\lambda =& - 3 k \Big(\lviIm + \lviiIm + (\lviIm - \lviiIm)c_{2\beta} + \lvIm s_{2\beta}\Big)^2,
\end{align}
\end{subequations}
where $k\equiv(4\pi)^{-2}$. The SM top-Yukawa coupling, $y_t$, is given in terms of the cTHDM top-Yukawa couplings by
\begin{align}\label{ytSMtoTHDM}
y_t(M_{H^\pm}) = \big|h_t \sbe + h_t' \cbe\big|\left(1 - \frac{3}{8} k \big|h_t \cbe - h_t' \sbe\big|^2\right).
\end{align}
The expressions given in \Eqss{lambdaSMtoTHDM}{ytSMtoTHDM} are only valid if the matching scale between the SM and the cTHDM is set equal to $M_{H^\pm}$.


\subsection{Matching the cTHDM to the MSSM}
\label{TDHMtoMSSM}

In addition to the calculation presented in \cite{Murphy:2019qpm}, we take into account all one-loop EWino contributions as well as the \order{\alt\als} correction for the matching of the THDM Higgs self-couplings. All threshold corrections between the cTHDM and the MSSM are derived in the limit of all sfermion masses, the gluino mass, and the matching scale $Q$ set equal to \msusy. Moreover, $M_1 = M_2 = \mu$ is assumed.


\subsubsection*{One-loop EWino contribution}

The one-loop EWino contributions to the matching of THDM Higgs self-couplings are given by
\begin{subequations}
\begin{align}
\Delta_\text{EWino}\li &= - \frac{1}{12}k\bigg(3 g^4 (8 + 7 \ln|\mf|^2) + 6 g^2 g_y^2 + 3 g_y^4 (2 + \ln|\mf|^2) \nonumber\\
&\hspace{2.0cm} + 4 g^2 g_y^2 \cos(\phi_{M_1} - \phi_{M_2})\bigg), \\
\Delta_\text{EWino}\lii &= - \frac{1}{12}k\bigg(3 g^4 (8 + 7 \ln|\mf|^2) + 6 g^2 g_y^2 + 3 g_y^4 (2 + \ln|\mf|^2) \nonumber\\
&\hspace{2.0cm} + 4 g^2 g_y^2 \cos(\phi_{M_1} - \phi_{M_2})\bigg), \\
\Delta_\text{EWino}\liii &= - \frac{1}{12}k\bigg(3 g^4 (8 + 7 \ln|\mf|^2) + 2 g^2 g_y^2 (4 - 3\ln|\mf|^2) + g_y^4 (8 + 9\ln|\mf|^2) \nonumber\\
&\hspace{2.0cm} - 4 g^2 g_y^2 \cos(\phi_{M_1} - \phi_{M_2})\bigg), \\
\Delta_\text{EWino}\liv &= - \frac{1}{6}k\bigg(3 g^4 (1 - \ln|\mf|^2)  + g^2 g_y^2  (7 + 15 \ln|\mf|^2) \nonumber\\
&\hspace{1.9cm} + 2 g_y^4 + 4 g^2 g_y^2 \cos(\phi_{M_1} - \phi_{M_2})\bigg), \\
\Delta_\text{EWino}\lv &= - \frac{1}{6}k e^{2i(\phi_{M_1} + \phi_{M_2})}\bigg(3 g^4 + 2 g^2 g_y^2 e^{2 i (\phi_\mu - \phi_{M_1} - \phi_{M_2})} + g_y^4\bigg), \\
\Delta_\text{EWino}\lvi &= - \frac{1}{3}k e^{i\phi_\mu}\bigg(3 g^4 e^{i\phi_{M_2}}+ g^2 g_y^2 (e^{i\phi_{M_1}}+e^{i\phi_{M_2}}) + g_y^4 e^{i\phi_{M_1}}\bigg), \\
\Delta_\text{EWino}\lvii &= - \frac{1}{3}k e^{i\phi_\mu}\bigg(3 g^4 e^{i\phi_{M_2}}+ g^2 g_y^2 (e^{i\phi_{M_1}}+e^{i\phi_{M_2}}) + g_y^4 e^{i\phi_{M_1}}\bigg),
\end{align}
\end{subequations}
where $\pMi$, $\pMii$ and $\pMue$ are the phases of the SUSY soft-breaking masses $M_1$, $M_2$ and $\mu$, respectively. $g$ is the $SU(2)_L$ gauge coupling; $g_y$, the $U(1)_Y$ gauge coupling. In the limit of vanishing phases, these expressions agree with the results given in~\cite{Gorbahn:2009pp,Bahl:2018jom}.

The different contributions to the one-loop threshold corrections of the top-Yukawa couplings can be categorized in the following way,
\begin{subequations}
\begin{align}
h_t^{\text{THDM}}(\msusy) =& h_t \left\{1 + k \left(\Delta_{\DR\leftrightarrow\MS} h_t + \Delta_{\tilde g} h_t + \Delta_{\tilde q} h_t + \Delta_{\tilde\chi} h_t \right)\right\}, \\
(h'_t)^{\text{THDM}}(\msusy) =& h_t k \left(\Delta_{\DR\leftrightarrow\MS} h'_t + \Delta_{\tilde g} h'_t + \Delta_{\tilde q} h'_t + \Delta_{\tilde\chi} h'_t \right),
\end{align}
\end{subequations}
where here $h_t$ is the the MSSM top-Yukawa coupling. The contributions arising from the change of the regularization scheme, $\Delta_{\DR\leftrightarrow\MS}$, have been derived in \cite{Bahl:2018jom} and are not modified in the presence of complex phases. The contributions from diagrams involving gluinos, $\Delta_{\tilde g}$, can be found in~\cite{Murphy:2019qpm}. In the same work, also the contributions from diagrams involving at least one squark (but no gluino), $\Delta_{\tilde q}$, have been derived in the limit of vanishing electroweak gauge couplings. The full expressions, including electroweak gauge couplings, are given by
\begin{subequations}
\begin{align}
\Delta_{\tilde q} h_t ={}& h_t^2\Big(\mathcal{F}_5(|\mf|)-\frac{1}{4}|\at|^2\Big)+ g^2 \mathcal{F}_1(|\mf|) \nonumber\\
&+ g_y^2 \left(\frac{17}{54}\mathcal{F}_5(|\mf|) - \frac{1}{9}|\at| |\mf| e^{i(\phi_{A_t} - \phi_{M_1})}\mathcal{G}_1(|\mf|) - \frac{1}{4}\mathcal{G}_2(|\mf|)\right),\\
\Delta_{\tilde q} h'_t ={}& \frac{1}{4} h_t^2|\at| |\mf| e^{-i(\phi_{A_t}+\phi_{\mu})} + g^2 e^{-i(\phi_{M_2}+\phi_\mu)}\mathcal{F}_2(|\mf|)+ g_y^2 e^{-i(\phi_{M_1}+\phi_\mu)}\mathcal{F}_4(|\mf|).
\end{align}
\end{subequations}
The loop-functions $\mathcal{G}_i$ are given by
\begin{subequations}
\begin{align}
\mathcal{G}_1(|\mf|) &= \frac{2}{(1-|\mf|^2)^2}\left(1 - |\mf|^2 + |\mf|^2 \ln |\mf|^2\right), \\
\mathcal{G}_2(|\mf|) &= \frac{2}{(1-|\mf|^2)^2}\left(-1 + |\mf|^2 - |\mf|^2 (2 - |\mf|^2) \ln |\mf|^2\right)
\end{align}
\end{subequations}
with the limiting values
\begin{align}
\mathcal{G}_1(0) &= 2, \hspace{1cm} \mathcal{G}_1(1) = 1, \hspace{1cm} \mathcal{G}_2(0) = -2, \hspace{1cm} \mathcal{G}_2(1) = 1.
\end{align}
The loop-functions $\mathcal{F}_{i}$ were defined in App.~A.3 of Ref.~\cite{Bahl:2018jom} and are given by
\begin{subequations}
\begin{align}
\mathcal{F}_1(|\mf|)=&\frac{3}{16(1-|\mf|^2)^2}\Big[7-4|\mf|^2-3|\mf|^4+2|\mf|^2(8-3|\mf|^2)\ln|\mf|^2\Big],\\
\mathcal{F}_2(|\mf|)=&\frac{3|\mf|^2}{2(1-|\mf|^2)^2}\Big[1-|\mf|^2+\ln|\mf|^2\Big],\\
\mathcal{F}_3(|\mf|)=&\frac{1}{144(1-|\mf|^2)^2}\Big[(55-32|\at||\mf|+51|\mf|^2)(1-|\mf|^2)\nonumber\\
&\hspace{2.5cm} + 2|\mf|^2(72-16|\at||\mf|-19|\mf|^2)\ln|\mf|^2\Big],\\
\mathcal{F}_4(|\mf|)=&\frac{|\mf|^2}{18(1-|\mf|^2)^2}\Big[13(1-|\mf|^2)+(9+4|\mf|^2)\ln|\mf|^2\Big],\\
\mathcal{F}_5(|\mf|)=&\frac{3}{8(1-|\mf|^2)^2}\Big[-1+4|\mf|^2-3|\mf|^4+2|\mf|^4\ln|\mf|^2\Big].
\end{align}
\end{subequations}
Using the identity
\begin{align}
\frac{17}{54}\mathcal{F}_5(|\mf|) - \frac{1}{9}|\at| |\mf| \mathcal{G}_1(|\mf|) - \frac{1}{4}\mathcal{G}_2(|\mf|) = \mathcal{F}_3(|\mf|),
\end{align}
the expressions given in \cite{Bahl:2018jom} are recovered if all phases are set to zero.

The contributions from diagrams involving only gauginos or Higgsinos is given by
\begin{subequations}
\begin{align}
\Delta_{\tilde\chi} h_t ={}& \frac{1}{12} (3 g^2 + g_y^2)(1 + 3\ln|\mf|^2),\\
\Delta_{\tilde\chi} h'_t ={}& \frac{1}{12} e^{-i \phi_\mu}(3 g^2 e^{-i\phi_{M_2}} + g_y^2 e^{-i\phi_{M_1}}).
\end{align}
\end{subequations}
Since no vertex diagrams involving only gauginos or Higgsinos exist, these corrections arise purely from external-leg contributions of the Higgs boson.


\subsubsection*{Two-loop \order{\alt\als} contribution}

The \order{\alt\als} contributions to the matching of THDM Higgs self-couplings are given by
\begin{subequations}
\begin{alignat}{2}
&\Delta_{\alt\als}\li &&= -\frac{4}{3}k^2 g_3^2 h_t^4 |\mf|^4, \\
&\Delta_{\alt\als}\lii &&= \frac{16}{3} k^2 g_3^2 h_t^4\bigg(-6|\at|\cos(\phi_{A_t} - \phi_{M_3})+3|\at|^2 \nonumber\\
& && \hspace{2.5cm} + |\at|^3\cos(\phi_{A_t}-\phi_{M_3}) - \frac{1}{4}|\at|^4\bigg), \\
&\Delta_{\alt\als}(\liii+\liv+\lvRe) &&= \frac{4}{3} k^2 g_3^2 h_t^4|\mf|^2\bigg(6 + 2|\at|\big(2\cos(\phi_{A_t}-\phi_{M_3})+\cos(\phi_{A_t}+\phi_{M_3}+2\phi_\mu)\big) \nonumber\\
& && \hspace{2.9cm} - |\at|^2\big(2+\cos(2\phi_{A_t}+2\phi_\mu)\big)\bigg), \\
&\Delta_{\alt\als}\lviRe &&= \frac{4}{3}k^2 g_3^2 h_t^4|\mf|^3\bigg(-\cos(\phi_{M_3}+\phi_\mu) + |\at|\cos(\phi_{A_t}+\phi_\mu)\bigg), \\
&\Delta_{\alt\als}\lviiRe &&= \frac{4}{3} k^2 g_3^2 h_t^4|\mf|\bigg(6\cos(\phi_{M_3}+\phi_\mu) - 6|\at|\cos(\phi_{A_t}+\phi_\mu) \nonumber\\
& && \hspace{2.8cm} - |\at|^2\big(2\cos(\phi_{M_3}+\phi_\mu)+\cos(2\phi_{A_t}-\phi_{M_3}+\phi_\mu)\big) \nonumber\\
& && \hspace{2.8cm} + |\at|^3\cos(\phi_{A_t}+\phi_\mu)\bigg).
\end{alignat}
\end{subequations}
These expressions are obtained from the \order{\alt\als} threshold correction between the SM and the MSSM~\cite{Bahl:2020tuq} using the strategy proposed in \cite{Lee:2015uza}. Since only the real parts of the $\lambda_i$'s enter the tree-level matching condition between the SM and the THDM (see \Eq{eq:SMtoTHDMtree}), it does not allow to obtain information about the imaginary parts of the Higgs self-couplings (we chose to set these contributions to zero). The complete \order{\alt\als} threshold correction have recently been published in~\cite{Bahl:2020jaq} but are not taken into account in this work. We expect the differences to the expressions presented here to have only a very small numerical effect.

Similar results have been obtained in~\cite{Carena:2015uoe}. In there, the stop soft-SUSY-breaking parameters have, however, been defined in the \MS scheme (see~\cite{Draper:2013oza} for more details).